\documentclass[twocolumn]{aastex631}

\usepackage{url}

\shorttitle{XRISM reveals low non-thermal pressure in Abell 2029}
\shortauthors{XRISM Collaboration}

\begin{document}

\title{XRISM reveals low non-thermal pressure in the core of the hot, relaxed galaxy cluster Abell 2029}

\author{XRISM Collaboration}%\altaffilmark{1}%
\affiliation{Corresponding Author: Eric D. Miller, milleric@mit.edu}
%\altaffiltext{1}{A-Address of Institute}
%\email{email address}

\author[0000-0003-4721-034X]{Marc Audard}
\affiliation{Department of Astronomy, University of Geneva, Versoix CH-1290, Switzerland} %1

\author{Hisamitsu Awaki}
\affiliation{Department of Physics, Ehime University, Ehime 790-8577, Japan} %2

\author[0000-0002-1118-8470]{Ralf Ballhausen}
\affiliation{Department of Astronomy, University of Maryland, College Park, MD 20742, USA} %3
\affiliation{NASA / Goddard Space Flight Center, Greenbelt, MD 20771, USA}
\affiliation{Center for Research and Exploration in Space Science and Technology, NASA / GSFC (CRESST II), Greenbelt, MD 20771, USA}

\author[0000-0003-0890-4920]{Aya Bamba}
\affiliation{Department of Physics, University of Tokyo, Tokyo 113-0033, Japan} %4

\author[0000-0001-9735-4873]{Ehud Behar}
\affiliation{Department of Physics, Technion, Technion City, Haifa 3200003, Israel} %5

\author[0000-0003-2704-599X]{Rozenn Boissay-Malaquin}
\affiliation{Center for Space Sciences and Technology, University of Maryland, Baltimore County (UMBC), Baltimore, MD, 21250 USA}
\affiliation{NASA / Goddard Space Flight Center, Greenbelt, MD 20771, USA}
\affiliation{Center for Research and Exploration in Space Science and Technology, NASA / GSFC (CRESST II), Greenbelt, MD 20771, USA}

\author[0000-0003-2663-1954]{Laura Brenneman}
\affiliation{Center for Astrophysics | Harvard-Smithsonian, Cambridge, MA 02138, USA} %7

\author[0000-0001-6338-9445]{Gregory V.\ Brown}
\affiliation{Lawrence Livermore National Laboratory, Livermore, CA 94550, USA} %8

\author[0000-0002-5466-3817]{Lia Corrales}
\affiliation{Department of Astronomy, University of Michigan, Ann Arbor, MI 48109, USA} %9

% ...

\begin{abstract}
     We present XRISM Resolve observations of the core of the hot, relaxed galaxy cluster Abell 2029. We find that the line-of-sight bulk velocity of the intracluster medium (ICM) within the central 180 kpc is at rest with respect to the Brightest Cluster Galaxy, with a 3-$\sigma$ upper limit of  $|v_{bulk}| < 100$ km\,s$^{-1}$. We robustly measure the field-integrated ICM velocity dispersion to be $\sigma_v = 169\pm10$ km\,s$^{-1}$, obtaining similar results for both single-temperature and two-temperature plasma models to account for the cluster cool core. This result, if ascribed to isotropic turbulence, implies a subsonic ICM with Mach number $\mathcal{M}_{3D} = 0.22$ and a non-thermal pressure fraction of 2.6$\pm$0.3\%. The turbulent velocity is similar to what was measured in the core of the Perseus cluster by Hitomi, but here in a more massive cluster with an ICM temperature of 7 keV, the limit on non-thermal pressure fraction is even more stringent. Our result is consistent with expectations from simulations of relaxed clusters, but it is on the low end of the predicted distribution, indicating that Abell 2029 is an exceptionally relaxed cluster with no significant impacts from either a recent minor merger or AGN activity.
\end{abstract}

\section{Introduction}

In the hierarchical structure formation paradigm, galaxy clusters are thought to grow over cosmic time through successive mergers and accretion of material from the cosmic web \citep[e.g.,][]{2008A&A...482L..29W,2015Natur.528..105E,2021A&A...647A...2R,2022ApJ...935L..23S}. The kinetic energy injected by successive merging events is released into the hot intracluster medium (ICM) and virialized, eventually turning into heat \citep{Kravtsov2012}. However, the sound crossing time in the medium can be as large as a Gyr, such that the thermalization process is expected to be slow. At the present day, a substantial fraction of the injected energy should remain in the form of kinetic motions, either in large-scale bulk motions or in an isotropic turbulent cascade \citep[e.g.][]{Lau2009,Vazza2009,Nelson2014,Biffi2016}. The remaining bulk and turbulent motions should generate shifts and broadening of the prominent X-ray plasma emission lines, which can then be used to assess the fraction of non-thermal energy within the ICM, estimate the thermalization timescale of the merging media, and constrain important plasma parameters such as its viscosity and thermal conduction.

Direct observational constraints on the non-thermal pressure fraction in galaxy clusters are scarce. In general, the spectral resolution of the CCD detectors on observatories like Chandra, XMM-Newton, and Suzaku is insufficient to measure Doppler shifts and broadening of the X-ray emission lines \citep[for a review, see][]{Simionescu2019}. Suzaku provided a few hints of large ICM bulk motions in a handful of cluster \citep{Tamura2011,Ota16}. More recently, a novel technique using instrumental fluorescence lines to correct the XMM-Newton EPIC pn energy scale has enabled more precise measurements of bulk motions in a number of clusters, allowing the first comparison of ICM kinematics in relaxed and merging clusters \citep{Sanders2020,Gatuzz_2022a,Gatuzz_2022b,Gatuzz_2023}. The Hitomi X-ray observatory provided the first direct measurement of the line-of-sight velocity broadening in the heart of the Perseus cluster \citep{Hitomi2016}.  This result suggests that in regions strongly affected by AGN feedback, the velocities in the core are subsonic and provide turbulent pressure support $\sim$4\% of the total pressure.  This result on a single cluster has waited eight years for confirmation in other clusters.

Abell 2029 is an ideal target to follow up the initial Hitomi result. It is a massive, hot galaxy cluster at $z=0.0787$ \citep{Sohnetal2019a} that exhibits a regular X-ray morphology, as evidenced by the surface brightness concentration and by the stability of the X-ray centroid in increasingly smaller apertures \citep{Eckert2022}. This firmly classifies the cluster as dynamically relaxed \citep{Cassano2010,Rossetti2017,Andrade2017}. X-ray imaging spectroscopy of the cluster by Chandra (see Figure \ref{fig:chandra}) indicates a mean atmospheric temperature of $7.5\pm 0.1$\,keV within the inner several hundred kpc and a cooler, $3$ keV central region with a cooling flow of several hundred solar masses per year \citep{Lewisetal2002,PaternoMahleretal2013,Martz2020}. Despite this cooling flow, its central galaxy IC\,1101 exhibits none of the classical signatures of cooling, such as optical nebular emission or the blue continuum emission from massive young stars. \cite{Dullo2017} suggested a SMBH mass of $(4$--$10)\times10^{10}$ $M_{\odot}$ for IC\,1101, based on its large depleted core in stellar light. With this massive SMBH, \cite{Prasad2024} showed that Bondi accretion of the hot gas can help to self-regulate the cool core with little precipitation.

The central galaxy hosts PKS\,1508+059, a powerful, double-lobed and wide-angle radio source. In cooling cores, the jets and lobes which periodically emanate from powerful radio sources are often seen to inflate buoyant cavities in the hot, X-ray emitting atmospheres. However, the X-ray atmosphere of Abell 2029 contains no detectable radio bubbles \citep{Clarke2004,PaternoMahleretal2013,Martz2020}.  Unsharp masked Chandra X-ray images, constructed by subtracting a mean two-dimensional surface brightness profile, have revealed a spiral structure extending from the central galaxy and curving out 70 kpc to the south-west, 150 kpc to the north, and 100 kpc to the east (see Figure \ref{fig:chandra}).  This swirling structure indicates the galaxy and its atmosphere are ``sloshing" with respect to each other. These harmonic motions are normally thought to be induced by mergers, or weaker gravitational interactions with a passing halo \citep{MarkevitchVikhlinin2007,ZuHone2010,ZuHoneetal2018,2023ApJ...944..132S}.  The sloshing motion in Abell 2029 
%may have erased cavities previously formed by the central radio source, and 
could indicate significant relative motion between the central galaxy and atmosphere with an amplitude of several hundred kilometers per second \citep{PaternoMahleretal2013}.  

We present XRISM \citep{Tashiro2020_XRISM} Resolve \citep{Ishisaki2022_Resolve} observations of the inner 3$\times$3 arcmin (270$\times$270 kpc) of Abell 2029 that aim to directly measure the velocity disperson and line-of-sight bulk velocity of the hot, bright ICM core. These relatively shallow observations are part of a larger campaign by the XRISM science team to map the velocity structure out to $R_{2500}$ (7.5 arcmin $\sim$ 670 kpc); the much deeper outer pointings are undergoing analysis and will be presented in future work.

Throughout this paper, we assume a $\Lambda$CDM cosmology with $H_0$ = 70 km\,s$^{-1}$ Mpc$^{-1}$, $\Omega_m$ = 0.3, and $\Omega_{\Lambda}$ = 0.7. At the redshift of Abell 2029 ($z = 0.0787$), 1 arcmin = 89 kpc. All reported redshifts and velocities have been corrected to the Solar System barycenter. We use the proto-solar abundance table from \cite{Lodders09}. Unless otherwise specified, reported uncertainties are 1-$\sigma$.

\section{Observations and data reduction}
\label{sect:obs}

\subsection{X-ray data with XRISM Resolve}
\label{sect:xrismdata}

The center of Abell 2029 was observed with XRISM over two co-aligned observations on January 10 and 13, 2024 (OBSIDS 000149000 and 000151000). The Resolve data for both observations was reprocessed with the XRISM team's Build 8 ftools software, applying calibration from CalDB v8 and default screening as described in the XRISM team's analysis of N132D \citep{XRISM2024_N132D}. The screening resulted in cleaned exposure times of 12.4 and 25.1 ksec for the two observations, for a total of 37.5 ksec.

\begin{figure}
\centering
\includegraphics[width=3.2in]{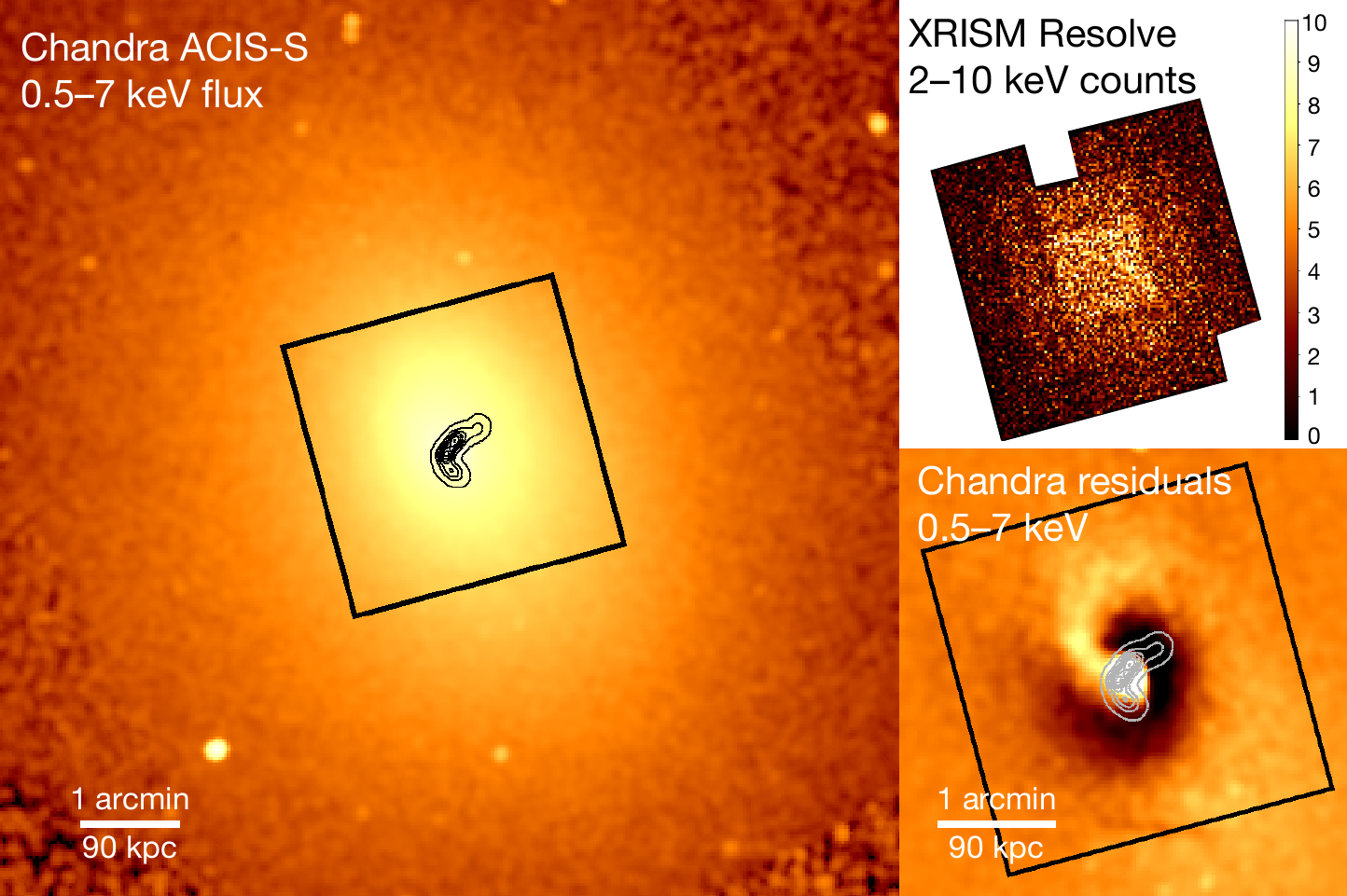}
\caption{(left) Chandra image of Abell 2029 overlaid with the 1.4 GHz VLA radio contours and the Resolve field of view. (top right) Resolve counts image from the two observations combined. Locations of individual photons are randomized within each pixel in the pipeline processing. (bottom right) Residuals after subtracting a 2-D model from the Chandra image, clearly showing the sloshing spiral that extends throughout most of the Resolve field. The Chandra data used were taken in 2004.}
\label{fig:chandra}
\end{figure}

The drift of the Resolve gain over an observation is tracked with occasional illumination of the focal plane with an $^{55}$Fe source mounted on the filter wheel, typically during Earth occultation. The first observation received five such fiducials, with an ADR recycle in the middle of the observation, and the second observation received six.  The flat-field-averaged energy scale uncertainty after gain reconstruction is $\leq$0.2 eV in the 5.4--8 keV energy band \citep{Eckart2024,Porter2024}.  We checked this using the calibration pixel, which is outside of the exposed array and constantly illuminated with a collimated $^{55}$Fe radioactive source. We reconstructed the gain for this pixel in the standard way, using the $^{55}$Fe filter wheel fiducials, and then fit the Mn K$\alpha$ 5.9-keV calibration line during only the on-source times, essentially mimicking the observation of a celestial source with this pixel. This produced an energy shift of $-0.12\pm0.03$ eV, consistent with the uncertainty quoted above in this energy band and comparable to other Resolve observations \citep[e.g.,][]{XRISM2024_N132D}.

We extracted spectra from the full array for each observation, excluding pixel 27, which in several observations has shown unexpected gain jumps that are not captured by the $^{55}$Fe fiducial cadence. Only high-resolution primary (`Hp' or \texttt{ITYPE=0}) events were included, as they account for more than 99\% of the 2--10 keV events in each observation. This fraction ignores low-resolution secondaries (`Ls' or \texttt{ITYPE=4}), which arise almost exclusively from instrumental effects at these low count rates. An image of the screened events is shown in Figure \ref{fig:chandra}. The spectra were binned using the optimal binning method of \citet{KaastraBleeker2016}, requiring at least one count per bin. A non-X-ray background (NXB) spectrum was extracted from a database of Resolve night-Earth data using \texttt{rslnxbgen} and weighting by the distribution of geomagnetic cut-off rigidity sampled during each observation. 

Responses were generated using the Build 8 ftools, the final pre-launch suite of XRISM software provided to the science team. The redistribution matrix file (RMF) was produced with an updated CalDB file for this build, \texttt{rmfparam\_20190101v006}, which as of this writing is the latest public CalDB file. We separated the electron loss continuum component as a more coarsely binned response file. The RMF was normalized by the Hp fraction in the 2--10 keV band, again ignoring Ls events, to account for the screened medium- and low-resolution events and ensure proper flux normalization. The anciliary response file (ARF) was constructed with \texttt{xaarfgen}, using an exposure-corrected 2--8 keV full-resolution Chandra image as the source model input. We also tested the analysis with a simpler point-source-model ARF, and while this alters the best-fit model normalization and derived flux, it has no effect on the reported velocity measurements within their statistical error. While we do not account explicitly for spatial-spectral mixing due to the instrumental PSF, we estimate from ray-tracing simulations with \texttt{xrtraytrace} that less than 10\% of the detected 2--10 keV X-ray emission originates from outside the Resolve field of view; 97\% originates within 2 arcmin (178 kpc) of the cluster center.

\subsection{Stellar velocity with MUSE}
\label{sect:muse}

We used the Multi-Unit Spectroscopic Explorer (MUSE) data of the central galaxy IC\,1101, to determine its velocity from stellar absorption lines. There were twelve 540-s MUSE WFM-AO observations from April to May 2018, with a total exposure time of 1.8 hours, under seeing conditions of 0.7$''$--1.4$''$ and airmass of 1.16--1.32. There were also six 110 sec observations on the night sky. We used the MUSE pipeline (version 2.9.0) with the ESO Recipe Execution Tool (EsoRex) to reduce the raw data, which provides a standard procedure to calibrate the individual exposures and combine them into a datacube. Further sky subtraction was performed with the Zurich Atmosphere Purge software (ZAP). We then extracted the spectrum within the central 5 kpc radius and examined the spectrum with pPXF. The system redshift derived from the stellar absorption lines is $z=0.0779\pm0.0001$, resulting in a BCG line-of-sight velocity uncertainty of $\pm30$ km\,s$^{-1}$. \cite{Sohnetal2019a} derived a redshift of 0.0787 for the full cluster and a redshift of 0.0778 for the BCG IC\,1101, which is consistent with our result.

\section{Analysis and discussion}
\label{sect:analysis}

\subsection{X-ray spectral fitting and modeling}
\label{sect:specfit}

\begin{table}
\caption{Best-fit parameters for the central region of Abell 2029 in the 2--10 keV energy band.\label{tab:best_params_center}}   
\begin{center}
\setlength{\tabcolsep}{6pt}
\begin{tabular}{lcc}
Parameter & {\tt AtomDB v3.0.9} & {\tt SPEXACT v3.07.00}\\ 
\hline
\hline
\multicolumn{3}{l}{1T model---{\tt TBabs$\star$bapec}} \\
\hline
  $kT$ (keV) & 6.83$_{-0.13}^{+0.14}$ & 6.88$_{-0.13}^{+0.13}$\\
  Abundance ($Z_{\odot}$) & 0.60$_{-0.02}^{+0.02}$ & 0.62$_{-0.02}^{+0.02}$\\
  Redshift\tablenotemark{\footnotesize a} & 0.07786$_{-0.00003}^{+0.00005}$ & 0.07786$_{-0.00003}^{+0.00004}$\\
  $v_{\rm bulk}$ (km\,s$^{-1}$)\tablenotemark{\footnotesize b} & $-10_{-7}^{+13}$ & $-9_{-8}^{+11}$ \\
  $\sigma_{v}$ (km\,s$^{-1}$) & 169$_{-10}^{+10}$ & 169$_{-10}^{+10}$\\
  $N$ (10$^{12}$ cm$^{-5}$)\tablenotemark{\footnotesize c} & 4.46$_{-0.08}^{+0.08}$ & 4.42$_{-0.08}^{+0.08}$\\
  C-stat/dof & 3892/3794 & 3894/3794\\
\hline
\multicolumn{3}{l}{2T model---{\tt TBabs$\star$(bapec+bapec)}} \\
\hline
 $kT_{1}$ (keV) & 5.02$_{-1.31}^{+0.34}$ & 4.22$_{-0.64}^{+0.52}$\\
 $kT_{2}$ (keV) & 10.85$_{-2.63}^{+1.36}$ & 9.91$_{-1.62}^{+1.58}$\\
 Abundance ($Z_{\odot}$) & 0.67$_{-0.04}^{+0.04}$ & 0.71$_{-0.04}^{+0.04}$\\
 Redshift\tablenotemark{\footnotesize a} & 0.07785$_{-0.00002}^{+0.00005}$ & 0.07785$_{-0.00003}^{+0.00004}$\\
 $v_{\rm bulk}$ (km\,s$^{-1}$)\tablenotemark{\footnotesize b} & $-13_{-7}^{+13}$ & $-11_{-7}^{+11}$ \\
 $\sigma_{v}$ (km\,s$^{-1}$) & 165$_{-10}^{+11}$ & 165$_{-10}^{+10}$\\
 $N_1$ (10$^{12}$ cm$^{-5}$)\tablenotemark{\footnotesize c} & 3.38$_{-1.6}^{+0.4}$ & 2.22$_{-1.0}^{+0.5}$\\
 $N_2$ (10$^{12}$ cm$^{-5}$)\tablenotemark{\footnotesize c} & 2.22$_{-0.4}^{+1.7}$ & 2.96$_{-0.5}^{+1.0}$\\
 C-stat/dof & 3882/3791 & 3882/3791\\
\hline
\end{tabular}
\vspace*{-\baselineskip}
\end{center}
\tablenotetext{a}{\footnotesize Redshift is with respect to the Solar System barycenter. Only statistical fitting errors are quoted and do not include instrumental systematic gain uncertainties.}
\tablenotetext{b}{\footnotesize Line-of-sight bulk velocity relative to the BCG redshift derived in Section \ref{sect:muse}, calculated as $\Delta v = c \Delta z / (1+z)$. Only statistical fitting errors are quoted and do not include the BCG redshift uncertainty ($\pm$30 km\,s$^{-1}$) or the instrumental systematic gain uncertainty.}
\tablenotetext{c}{\footnotesize The model normalization corresponds to a 3-arcmin-radius circular region centered on Abell 2029, since this was the model input to \texttt{xaarfgen}.}
\end{table}

We fit the full-field Resolve spectra using \texttt{Xspec v12.14.0h} \citep{1996ASPC..101...17A} and employing C-statistics \citep{1979ApJ...228..939C}. For plasma model calculations, we utilized the atomic databases from {\tt AtomDB v3.0.9} \citep{2012ApJ...756..128F} and {\tt SPEXACT v3.07.00} \citep{1996uxsa.conf..411K}. The \texttt{SPEX} continuum and line emission files were converted into \texttt{Xspec}-readable format, enabling us to directly compare results between two databases and ensuring a consistent approach across all underlying assumptions and fitting procedures. To model ICM plasma in both databases, we adopted a velocity-broadened collisional-equilibrium model (\texttt{bapec} in \texttt{Xspec}), allowing for a single redshift and velocity broadening in addition to the plasma thermal broadening. We applied both single-temperature (1T) and two-temperature (2T) models to explore whether the cool-core affected the velocity measurements. A multiplicative absorption model \citep[\texttt{TBabs} in \texttt{XSPEC},][]{2000ApJ...542..914W} was included to account for the line-of-sight Galactic neutral column, with $N_{\rm H}$ fixed at $3\times10^{20}$ cm$^{-2}$ \citep{2016A&A...594A.116H}.  All other fit parameters were allowed to vary, tied between the spectra from each observation, which were observed close enough in time that the heliocentric correction is essentially the same ($+$25.5 km\,s$^{-1}$ for the first observation, $+$26.0 km\,s$^{-1}$ for the second). For the 2T model, we tied the abundance, redshift, and velocity dispersion $\sigma_{v}$ between the two components. 

\begin{figure*}
\centering
\includegraphics[width=6.5in]{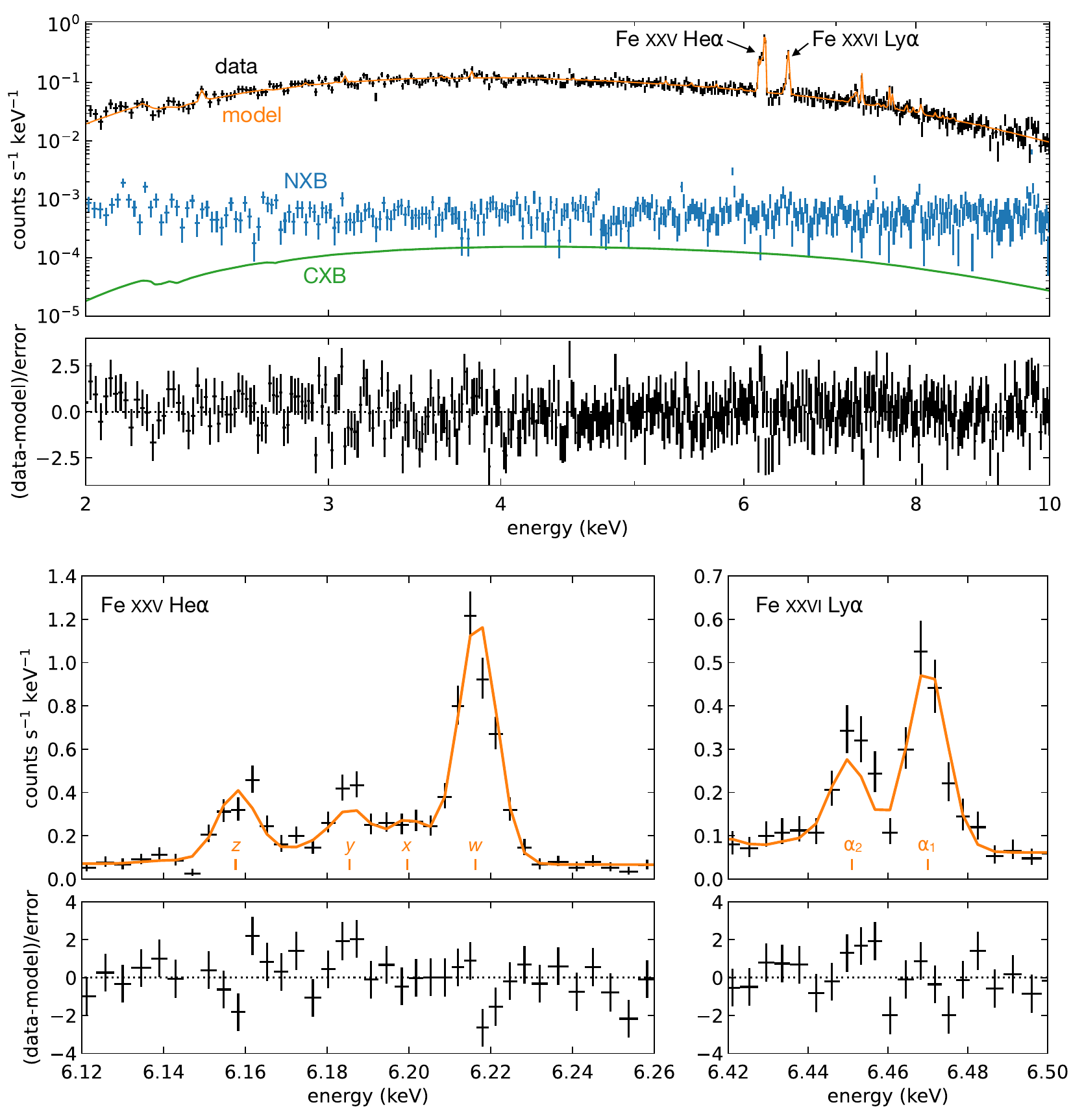}
\caption{(top) Broad-band Resolve spectrum of the center of Abell 2029, with the two observations summed for clarity. The best-fit 1T \texttt{AtomDB} model is shown in orange. The NXB extracted from observations of the dark limb of the Earth is shown in blue, and is at least one order of magnitude below the cluster emission at all energies. Similarly, the estimated unresolved CXB shown in green is well below the cluster emission. (bottom) Resolve spectrum zoomed into the strongest emission features of He-like (left) and H-like (right) Fe, as these lines dominate the velocity fit. The individual components of each line complex are distinguishable; in the He-like triplet, these are the forbidden ($z$), intercombination ($y,x$), and resonance ($w$) transitions. The fit residuals show no systematic features around these sharp emission lines, although there are some deviations that are discussed in the text.}
\label{fig:spec}
\end{figure*}

The best-fit model parameters are listed in Table \ref{tab:best_params_center}, and the combined spectrum with best-fit 1T \texttt{AtomDB} model is shown in Figure \ref{fig:spec}. While we report the values obtained from a broad-band 2--10 keV fit, fits within a narrow band around the He- and H-like Fe K$\alpha$ line complexes (5.5--7 keV) produced identical redshift and velocity broadening within the uncertainty. This indicates that these complexes dominate our leverage on ICM velocity, a fact also apparent from their visual prominence in the spectrum. We estimate the systematic instrumental uncertainty in $v_{\rm bulk}$ to be $\pm10$ km\,s$^{-1}$, based on the $0.2$ eV gain uncertainty described in Section \ref{sect:obs} for energies near 6 keV. The systematic uncertainty in the line-of-sight velocity dispersion is $\pm3$ km\,s$^{-1}$, based on propagating the 0.3 eV FWHM uncertainty of the underlying core instrumental line-spread function.

The total background is negligible compared to the observed source spectrum. The NXB spectral flux is at least one order of magnitude lower than that of Abell 2029 at all energies across the 2--10 keV band (blue points in Figure \ref{fig:spec}, left). We estimated the contribution from unresolved background AGN (the Cosmic X-ray Background or CXB; green line in Figure \ref{fig:spec}, left) using observations from Suzaku \citep{Bautz2009}, which has a similar PSF to XRISM. This contribution is even lower than the NXB. We have therefore excluded the background from our spectral fits.

Resonant scattering could suppress the flux of the strongest \textit{w} (resonance) line in the He-like Fe K$\alpha$ complex. Following the method used for the Hitomi observation of Perseus \citep{Hitomi3_2018}, we checked that the effect is not detectable in the considered region with the current exposure time by repeating the modeling with and without this line. There was no significant effect on the velocity measurements. We also note that the observed flux of the $y$ intercombination line within the He-like Fe K$\alpha$ line complex is considerably higher than the model predicts. This does not affect the velocity measurements. We comment on both of these effects in the context of future work in the next section.

\subsection{Velocity structure of the hot ICM}
\label{sect:velocity}

In Table~\ref{tab:best_params_center}, we estimated the line-of-sight bulk velocity of the gas relative to the BCG as $-11^{+12}_{-7}~{\rm km\,s^{-1}}$, averaging over the similar results from the four spectral fits. Taking the quadrature sum of the statistical error, the BCG redshift uncertainty ($\pm30$ km\,s$^{-1}$), and the instrumental uncertainty ($\pm10$ km\,s$^{-1}$), the 3-$\sigma$ upper limit of bulk velocity becomes $|v| < 100~{\rm km\,s^{-1}}$. This indicates no significant line-of-sight velocity difference between the BCG and the ICM. While this result is consistent within the error range of previous Suzaku XIS measurements of the central region \citep{Ota16}, the upper limit has become an order of magnitude stricter. Therefore, the bulk motion is neglected in the following pressure calculation.

The central galaxy hosts the powerful wide-angle tail radio source PKS 1508+059 \citep{Lewisetal2002,PaternoMahleretal2013,Martz2020}, extending 40 kpc into the atmosphere (see Figure \ref{fig:chandra}).  Despite its high radio power of $P_{1.4} = 10^{41}$ erg\,s$^{-1}$, X-ray cavities are absent.  Cavities that may have formed earlier in the development of the radio source were possibly destroyed by the sloshing spiral or by hydrodynamical instabilities \citep{PaternoMahleretal2013}.  Applying the \citet{Cavagnolo2010} scaling relation between synchrotron power and mechanical power, we find a total mechanical power of $\sim 3 \times 10^{44}$ erg\,s$^{-1}$.  Assuming the radio source has remained active for $10^7$ yr to $10^8$ yr, it would have deposited $10^{59-60}$ erg into the atmosphere.  The atmospheric gas mass encompassed by the Resolve pointing is $4\times10^{12}\,M_{\odot}$ based on a Chandra X-ray imaging analysis.  The total kinetic energy of the atmosphere implied by $\sigma_v \approx 170$ km\,s$^{-1}$ is then $3.5\times10^{60}$ erg.  Therefore, the central radio jet could have contributed a few to several tens of percent of the atmospheric kinetic energy within the pointing.  This assumes no other sources of extra-thermal motions (e.g., turbulence).  A full analysis that includes a sub-grid estimate of the velocity dispersion near the jet and a comparison to the off-nuclear pointing will appear in a future paper.

The velocity dispersion of $169\pm10$ km\,s$^{-1}$ reported here for the central region of Abell 2029 is low in comparison with the sound speed in the system. For a mean temperature of 6.8 keV, the sound speed in the medium is $c_s=\left(\gamma k_BT/\mu m_p \right)^{1/2} = 1340$ km\,s$^{-1}$, with $\gamma=5/3$ the adiabatic index and $\mu=0.61$ the mean molecular weight in a fully ionized plasma. If the velocity dispersion determined by Resolve is entirely ascribed to isotropic turbulence, our measurement implies a turbulent Mach number $\mathcal{M}_{3D}=\sqrt{3}\sigma_v/c_s = 0.22$; turbulent motions are highly subsonic. The resulting non-thermal (NT) pressure fraction becomes \citep{Eckert2019}
\begin{equation}
     \frac{P_{NT}}{P_{\rm tot}} = \frac{\mathcal{M}_{3D}^2}{\mathcal{M}_{3D}^2+3/\gamma} = 0.026\pm0.003 ,
\end{equation}
where we assume that all NT pressure comes from turbulence and ignore contributions from other non-thermal components such as magnetic fields and cosmic rays \citep{MiniatiBeresnyak2016,ettori2022}. The uncertainty in the NT pressure fraction is calculated from the statistical uncertainties in the best-fit velocity dispersion and temperature of the 1T \texttt{AtomDB} model.

In comparison with the core of the Perseus cluster \citep{Hitomi2016,Hitomi2_2018,Hitomi3_2018}, we measure a similar velocity dispersion in a more massive cluster, such that the limit on the NT pressure fraction is even more stringent. The absence of obvious AGN feedback features in Abell 2029 may explain the lower NT pressure fraction in this cluster compared to Perseus, although the observations probe very different physical scales in the two clusters.

Such a low level of turbulence-induced NT pressure is consistent with estimates based on indirect methods, including gas pressure, density, and temperature power spectra \citep[e.g.,][]{schuecker2004,Zhuravleva2014,Zhuravleva2018,Heinrich2024,Dupourque2023,Dupourque2024,Lovisari2024}, universal ICM gas fraction \citep{Eckert2019,ettori2022,Wicker2023}, and three-dimensional multi-probe reconstructions \citep{Sayers2021,Umetsu2015,Umetsu2022}. All of these methods imply a low level of NT pressure in cluster cores ($<$10\%), in agreement with direct high-resolution X-ray measurements in Perseus and now Abell 2029. Studies of the NT pressure fraction in cosmological simulations predict that it should increase with radius, from $\sim$10\% in cluster cores to $\sim$30\% at the virial radius \citep[e.g.,][]{Lau2009,Vazza2009,Nelson2014,Biffi2016,Vazza2018,Angelinelli2020,Gianfagna2023}. However, there exists some level of discrepancy on the impact of NT pressure in cluster cores between studies using different hydrodynamical solvers (adaptive mesh refinement or smoothed particle hydrodynamics) and implementing different baryonic physics (AGN feedback, cooling, and star formation). Additionally, it is unclear whether the calculation of the NT pressure fraction in simulations should be done using the entire velocity field or random gas motions only \citep{Vazza2018,Angelinelli2020}.

Recently, \citet{Truong2024} studied the level of NT pressure in the cores of Perseus-like clusters in the {\tt TNG-Cluster} simulation set \citep{Nelson2024}, which provides resimulations of a sample of $>$300 massive clusters with the TNG model \citep{Pillepich2018}. The median NT pressure fraction inferred from mock XRISM data is $\sim$8\%, with some systems exhibiting NT pressure fractions as large as $\sim$15\%. Our measurement of Abell 2029 occupies the lower end of the \citet{Truong2024} distribution in a system that is more massive than Perseus. 

We note that Abell 2029 is an exceptionally relaxed-looking cluster away from the core. Even within the core, the sloshing structure is very low in contrast compared to other cluster cool cores, and simulations show that sloshing should last for many Gyr after the disturbing event. In Abell 2029, it appears that event occurred several Gyr ago. The AGN is currently bright in the radio, but there is no evidence in the X-rays of any AGN-driven mechanical disturbance. These characteristics, combined with the low NT pressure support inferred from the low XRISM velocity dispersion, suggest the cluster may currently be in a quiescent phase of AGN feedback (i.e., no active energy pumping into the ICM at this stage), with any recent minor merger only having enough time to mildly perturb the gas velocity. 

Two results from our focused study require future work to fully understand. First, the lack of measurable resonant scattering is at first glance puzzling, given the low velocity dispersion and the detection of resonant scattering in Perseus \citep{Hitomi3_2018}. Indeed, using the model presented by \cite{Zhuravleva2013b}, we predict that the optical depth at the center of the Fe$\,${\small\rmfamily{XXV}} He$\alpha\,w$ component should be higher in Abell 2029 ($\tau\sim3$) than in Perseus ($\tau\sim2$). However, we estimate that more than 150 ks is required to significantly detect the change in line strength at the current sensitivity of Resolve. Our 37-ks observation is simply not deep enough, compared to the 244-ks Hitomi observation of the Perseus core. A deeper XRISM observation approved for AO1 may robustly detect resonant scattering in Abell 2029. Second, several transitions are brighter in Figure \ref{fig:spec} than predicted by either the 1- or 2-temperature models, notably the Fe$\,${\small\rmfamily{XXV}} He$\alpha\,y$ intercombination line and the Fe$\,${\small\rmfamily{XXVI}} Ly$\alpha_2$ resonance component. The former is not due to resonant scattering as it persists even when the resonance ($w$) line is excluded from the fit. It could result from multi-temperature structure, and this will be explored in a future paper. The residuals around the Ly$\alpha_2$ are unexplained at this time. Future work will compare this feature in Abell 2029 to that in other thermal plasmas observed with Resolve to understand if there is a systematic effect.

Our results here from an observation of modest depth demonstrate the power of XRISM to measure line-of-sight velocity structure in galaxy clusters. Additional XRISM measurements in several other systems and out to larger radii in Abell 2029 are currently being analyzed or planned for observation. This larger dataset will allow us to compare measurements of the NT pressure fraction over a sizable cluster sample spanning a range of dynamical states, and to constrain the importance of NT pressure in the outer regions of clusters where AGN activity should be less important.

\vspace*{\baselineskip}
%\begin{acknowledgments}
We gratefully acknowledge the hard work over many years of all of the engineers and scientists who made the XRISM mission possible. Part of this work was performed under the auspices of the U.S. Department of Energy by Lawrence Livermore National Laboratory under Contract DE-AC52-07NA27344. The material is based upon work supported by NASA under award numbers 80GSFC21M0002 and 80GSFC24M0006. This work was supported by JSPS KAKENHI grant numbers JP22H00158, JP22H01268, JP22K03624, JP23H04899, JP21K13963, JP24K00638, JP24K17105, JP21K13958, JP21H01095, JP23K20850, JP24H00253, JP21K03615, JP24K00677, JP20K14491, JP23H00151, JP19K21884, JP20H01947, JP20KK0071, JP23K20239, JP24K00672, JP24K17104, JP24K17093, JP20K04009, JP21H04493, JP20H01946, JP23K13154, JP19K14762, JP20H05857, and JP23K03459. This work was supported by NASA grant numbers 80NSSC20K0733, 80NSSC18K0978, 80NSSC20K0883, 80NSSC20K0737, 80NSSC24K0678, 80NSSC18K1684, 80NSSC23K0650, and 80NNSC22K1922. LC acknowledges support from NSF award 2205918. CD acknowledges support from STFC through grant ST/T000244/1. LG acknowledges financial support from Canadian Space Agency grant 18XARMSTMA. MS acknowledges the support by the RIKEN Pioneering Project Evolution of Matter in the Universe (r-EMU) and Rikkyo University Special Fund for Research (Rikkyo SFR). AT and the present research are in part supported by the Kagoshima University postdoctoral research program (KU-DREAM). SY acknowledges support by the RIKEN SPDR Program. IZ acknowledges partial support from the Alfred P. Sloan Foundation through the Sloan Research Fellowship. SE acknowledges the financial contribution from the contracts Prin-MUR 2022 supported by Next Generation EU (M4.C2.1.1, n.20227RNLY3 {\it The concordance cosmological model: stress-tests with galaxy clusters}), ASI-INAF Athena 2019-27-HH.0, ``Attivit\`a di Studio per la comunit\`a scientifica di Astrofisica delle Alte Energie e Fisica Astroparticellare'' (Accordo Attuativo ASI-INAF n. 2017-14-H.0), and from the European Union’s Horizon 2020 Programme under the AHEAD2020 project (grant agreement n. 871158). LL acknowledges support from INAF grant 1.05.12.04.01. This work was supported by the JSPS Core-to-Core Program, JPJSCCA20220002. The material is based on work supported by the Strategic Research Center of Saitama University.
%\end{acknowledgments}

\end{document}